# Strict equivalence between Maxwell-Stefan and fast-mode theory for multicomponent polymer mixtures

*Olivier J.J. Ronsin[1] and Jens Harting[1,2*]*

1: Helmholtz Institute Erlangen-Nürnberg for Renewable Energy, Forschungszentrum Jülich, Fürther Straße 248, 90429 Nürnberg, Germany

2: Department of Applied Physics, Eindhoven University of Technology, P.O. box 513, 5600MB Eindhoven, The Netherlands

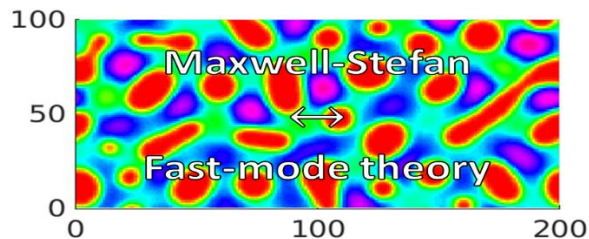

Table of contents: for Table of Contents use only

# Abstract

The applicability of theories describing the kinetic evolution of fluid mixtures depends on the underlying physical assumptions. The Maxwell-Stefan equations, widely used for miscible fluids, express forces depending on coupled fluxes. They need to be inverted to recover a Fickian form which is generally impossible analytically. Moreover, the concentration dependence of the diffusivities has to be modelled, e.g. by the multicomponent Darken equation. Cahn-Hilliard type equations are preferred for immiscible mixtures, whereby different assumptions on the coupling of fluxes lead to the slow-mode and fast-mode theories. For two components, these were derived from the Maxwell-Stefan theory in the past. Here, we prove that the fast-mode theory and the generalized Maxwell-Stefan theory together with the multicomponent Darken equation are strictly equivalent even for multicomponent systems with very different molecular sizes. Our findings allow to reduce the choice of a suitable theory to the most efficient algorithm for solving the underlying equations.

# Introduction

The kinetic evolution of fluid mixtures is important in many scientific fields ranging from energy technology to biology or nanotechnology. In reactive or membrane systems, the precise knowledge of diffusion-driven transport of the different chemical species is crucial to estimate the reaction or separation efficiency[1] [2]. In polymer solutions and blends, the fluids are often immiscible and various morphologies arise upon demixing and influence the properties of the final system. This is typically the case in solution-processed thin films used for example for coatings, membranes, organic electronics or organic photovoltaics [3] [4] [5] [6] [7].

On the one hand, the Maxwell-Stefan equations [1] [8] [9] are often used to study diffusion in miscible multi-component systems. In their original form, they relate the flux to the concentration gradients but later they have been extended to the generalized Maxwell-Stefan equations (GMS) [10] [11] [12] relating the flux to any driving force in the system. The GMS equations have been extended for fluids with very different molecular sizes by Curtiss [13] and later Fornasiero [14] with the hypothesis that all segments of a macromolecule are available for friction with molecules of the other species. On the other hand, for immiscible systems, one way of investigating the spinodal decomposition and subsequent ripening of the phases is to use Cahn-Hilliard type equations [15] [16]. These are classical diffusion equations, but with a particular driving force for the system evolution, namely the chemical potential gradient, which includes not only mixing terms but also surface tension terms in order to handle separated phases. Both Cahn Hilliard theory and Maxwell-Stefan theory aim at describing diffusion processes in the framework of classical mechanics and one can wonder about the relationship between them. In particular, it is questionable whether the Cahn-Hilliard equations could be understood as GMS equations with the chemical potential gradient taken as the driving force. With this condition, it seems that the physics is very similar, but the equations themselves are not formally equivalent. Moreover, in the framework of each theory, different assumptions can be made to describe the interactions between the fluid fluxes, leading to various particular models reviewed below.

One of the major difficulties with Maxwell-Stefan equations is that the driving force is written as a function of all involved fluxes, so that the coupled equations have to be inverted to compute the fluxes, which is in the general case not possible analytically. Additionally, the mobility coefficients appearing in the equations (the so-called Maxwell-Stefan diffusivities) are not easily accessible by experiments. Three main approaches have been proposed to determine the Maxwell-Stefan diffusivities[17]. First, Fickian mutual diffusion coefficients can be measured in a multicomponent mixture depending on the concentration. The Maxwell-Stefan diffusivities are subsequently calculated by inversion of the system of equations[18]. Second, they can also be extracted from molecular dynamics simulations [19]. Third, various phenomenological or theoretical models have been developed in order to obtain relationships between Maxwell-Stefan diffusivities and the (tracer) self-diffusion coefficients of each material, whereby the latter have to be determined through either experiments or computer simulations [17][20]. Darken and Vignes proposed in 1948 and 1966 expressions for the Maxwell-Stefan diffusivities depending on the self-diffusion coefficients for binary mixtures[21][22], and these expressions have been recently generalized to multicomponent systems from statistical mechanics considerations within the linear response theory[19]. Other phenomenological models exist, but the multicomponent Vignes and Darken model are the most popular and successful. Liu, Bardow and Vlugt also proposed to refine these models so that the Maxwell-Stefan diffusivities can be expressed as a function of the self-diffusion coefficients at infinite dilution only. The corresponding expressions are known as the "Vignes-LBV" equation [23] and the "Darken-LBV" equation [19]. They have been proven to match very well to MD simulations [19] and experimental results [24] in systems where the intermolecular interactions are weak. Furthermore, it is consistent with the main hypothesis underlying the derivation of the multicomponent Darken equation from statistical mechanics, namely neglecting cross correlations of velocities for different molecules.

Since the appearance of the original Cahn- Hilliard equations for a two-component system, different theories have been developed to properly describe the kinetics of the phase evolution for multicomponent systems as well. These equations have been implemented in time-resolved phase-field simulations which have been widely applied to investigate immiscible mixtures [25][26][27][28][29][30] even in drying systems [31][32][33][34][35]. One important difference between the various studies on the simulation of spinodal decomposition is the relationship between the flux and the driving forces. In Cahn's original assumption, the flux is simply proportional to the gradient of the generalized exchange chemical potential through a mobility coefficient. However, as a consequence of the Gibbs-Duhem relationship together with the incompressibility constraint, the fluxes of all materials are coupled together and the mobility coefficients depend on the volume fraction. The expression of the mobility coefficients is not trivial in systems containing more than two fluids and several theories have been proposed to derive correct expressions for the flux. At the atomic scale, diffusion is supposed to occur upon movement of the material atoms (or molecules) into free space or vacant sites in the material. These "vacancies" move in the opposite direction. In the so-called "slow mode theory", initially proposed by De Gennes and Brochard, the material fluxes are assumed to compensate themselves exactly in a reference frame fixed to the material atomic or molecular lattice (the "lattice" reference frame), meaning that the vacancy flux is equal to zero. The incompressibility condition is introduced with the help of an additional potential that can be interpreted as a chemical potential for the vacancies in the system[36][37]. Shortly after, Kramer proposed the so-called "fast-mode theory"[38]. In contrast to Brochard and De Gennes, he proposed that the flux of vacancies cannot be zero: it has to compensate the fluxes of all the materials in the lattice reference frame. However, the chemical potential of the vacancies is zero, i.e. the vacancy concentration is at equilibrium. Both theories lead to different expressions for the mobility coefficient, from which the mutual diffusion coefficient can be easily calculated in a two-component system. It turns out that for the "slow-mode theory", the mutual diffusion coefficient is controlled by the slowest component, while it is controlled by the fastest component in the "fast-mode theory". These considerations gave the names for both theories. Another important feature of the fast-mode theory is that in the lattice reference frame, the flux of both components is not balanced, so that it predicts a Kirkendall effect: inert markers positioned at an initial given interface seem to move in the laboratory frame as a result of diffusion[39][40]. In the slow-mode theory this is by construction not the case: since the vacancy flux is equal to zero, the atomic fluxes of two different materials across an interface have to be equal, as if diffusion was occurring by a direct exchange mechanism, in which atoms migrate by switching positions. Akcasu, Naegele and Klein later proposed a more general theory (ANK) bridging the two limiting cases of slow-mode and fast-mode for binary systems[41][42]. It should be noted that the slow-mode theory is described as an "incompressible" theory since the incompressibility condition is explicitly used to derive the equations. In contradiction to that, the fast-mode theory is often described as a theory for locally compressible systems, although the incompressibility condition is actually perfectly fulfilled once the fluxes are written in the volume reference frame. There have been numerous experimental studies aiming at checking the validity of these theories (see e.g. [38][43][44] and [45] for further references). It turns out that in most tested polymer systems the fastest component drives the kinetics, and the Kirkendall effect can also be observed. As a conclusion, the idea emerging from the literature is that the fast-mode theory might better match the experimental observations in comparison to the slow-mode theory.

Finally, the fast-mode theory on the one hand and the Maxwell-Stefan theory together with the Darken-LBV equation on the other hand seem to be two of the most successful theories for modeling the kinetic behavior of multicomponent systems. Both theories have a sound theoretical basis and are supported by different experimental studies. Apparently, they were developed and used in two distinct scientific communities and there are actually very few studies dealing with the relationship between them. Foley derived both fast- and slow-mode theory from the Maxwell-Stefan equations for binary systems assuming two different hypotheses for the friction coefficient between molecules of the different species[46]. Liu and co-workers also noticed that the multicomponent Darken equation for binary systems is clearly equivalent to the fast-mode equation [19].

However, to the best of our knowledge, the relationship between both theories for the general case of a mixture with any number of components has not been investigated. In the remainder of this paper we prove that the fast-mode theory and the generalized Maxwell-Stefan theory together with the multicomponent Darken equation are strictly equivalent even for multicomponent systems. This also holds for mixtures of fluids with very dissimilar molecular size.

# Kinetic equation for Maxwell-Stefan theory with multicomponent Darken diffusivities

## 1.   Generalized Maxwell-Stefan equations for molecules with dissimilar sizes

The generalized Maxwell-Stefan equations for molecules with dissimilar molar sizes can be written in the form

$$\sum_{j=1}^{n} \frac{1}{c_{tot} D_{ij}} \left( \frac{\varphi_j}{v_j} \boldsymbol{j}_i - \frac{\varphi_i}{v_i} \boldsymbol{j}_j \right) = -\frac{\varphi_i}{RT} \nabla \mu_{V,i}^{mix}, \qquad i = 1 \dots n,$$

(1)

where $c_{tot}$ is the total concentration and $\varphi_i$, $v_i$, $\boldsymbol{j}_i$, are the volume fractions, molar volumes and fluxes of the fluids $i$, respectively. The coefficients $D_{ij}$ are the symmetric Maxwell-Stefan diffusivities between fluid $i$ and fluid $j$. $R$ is the molar gas constant and $T$ the temperature. $\mu_{V,i}^{mix}$ is the exchange chemical potential defined as a volume variable, meaning that $\mu_{V,i}^{mix} = \mu_i^{mix}/v_i$, where $\mu_i^{mix}$ is the exchange chemical potential. Note that the driving forces have been restricted here to the chemical potential gradients, but thermal, gravitational and pressure terms can be taken into account without loss of generality of the present demonstration. Following Fornasiero [14], we define the molecules as chains of $N_i$ connected segments or beads interacting with other molecules upon elementary collisions, having a molar volume $v_0$ (which may typically be chosen as the smallest molar volume in the system) and assume that all beads of a molecule have the same frictional properties and equally contribute to the friction independently of the size of the molecule. With these assumptions and $D_{ij}{}^0$ being the Maxwell-Stefan diffusivities of the molecule beads, we have $v_i = N_i v_0$ and $D_{ij}{}^0 = N_i D_{ij}$, so that the GMS can be rewritten as

$$\sum_{j=1}^{n} \frac{1}{D_{ij}{}^0}\left(\varphi_j \boldsymbol{j}_i - \varphi_i \boldsymbol{j}_j\right) = -\frac{v_0 \varphi_i}{RT}\nabla \mu_{V,i}^{mix}, \qquad i = 1 \dots n.$$

(2)

Only *n-1* equations are independent and with the incompressibility hypothesis $\sum_{i=1}^{n} \boldsymbol{j}_i = 0$, this relationship can be formally inverted to match the more classical Fickian form [1] [47] [48]

$$\frac{\partial \varphi_i}{\partial t} = \nabla\left[-\frac{v_0}{RT}B^{-1}\varphi\nabla\mu_V^{mix}\right], \qquad i = 1 \dots n-1,$$

(3)

where $\nabla\mu_V^{mix}$ is the vector of the *n-1* first chemical potential gradients, $\varphi$ the vector of the *n-1* first volume fractions and $B$ the matrix defined by

$$\begin{cases} B_{ii} = \left( \dfrac{\varphi_i}{{D_{in}}^0} + \displaystyle\sum_{k=1, k \neq i}^{n} \dfrac{\varphi_k}{{D_{ik}}^0} \right) \\ B_{ij} = \varphi_i \left( \dfrac{1}{{D_{in}}^0} - \dfrac{1}{{D_{ij}}^0} \right) \end{cases}.$$

(4)

## 2. GMS with multicomponent Darken diffusivities

The multicomponent Darken model for Maxwell-Stefan diffusivities reads[19]

$$D_{ij} = \frac{D_{s,i} D_{s,j}}{D_{mix}} = D_{s,i} D_{s,j} \sum_{k=1}^{n} \frac{x_k}{D_{s,k}},$$

(5)

where $D_{s,i}$ is the (volume-fraction dependent) self-diffusion coefficient of the fluid $i$ with mole fraction $x_i$ and where $D_{mix}$ has been defined by the following equation:

$$\frac{1}{D_{mix}} = \sum_{k=1}^{n} \frac{x_k}{D_{s,k}},$$

(6)

This expression is derived from statistical mechanics in the following way [17] [19] [23]: in a multicomponent system, the Onsager mobility coefficients, which correspond to Maxwell-Stefan mutual diffusivities, are written with the help of the Green-Kubo formula depending on correlations between the velocities of all molecules in the system. These correlations contains three contributions, namely the auto-correlation function of the molecules, the cross-correlation function between different molecules of the same fluid, and the cross-correlation function between molecules of different fluids. Neglecting all the cross-correlations makes the Maxwell-Stefan diffusivities only depend on the auto-correlation functions. Thus, in other words on the self-diffusion coefficients. The calculation leads finally to the expression above. The multicomponent Darken has therefore a sound theoretical basis. However, it is supposed to be valid only for systems where the cross-correlations are small as compared to the auto-correlations. This corresponds to systems where molecules weakly interact with each other. As a consequence, this should be considered cautiously when dealing with hydrogen bonding or polar fluids, or with polymer solutions or blends in the entangled regime.

We now use this expression to find a simplified form for the $B$ matrix, Eq. (4). Since the model is written for segments of molecules which are set to have all the same size, the mole and volume fraction of these segments are equal so that the multicomponent Darken equation (5) becomes

$$D_{ij}{}^0 = \frac{D_{s,i}{}^0 D_{s,j}{}^0}{D_{mix}} = D_{s,i}{}^0 D_{s,j}{}^0 \sum_{k=1}^{n} \frac{\varphi_k}{D_{s,k}{}^0}.$$

(7)

The coefficients of the matrix $B$ can then be expressed as

$$B_{ii} = \left( \frac{\varphi_i}{D_{in}{}^0} + \sum_{k=1, k \neq i}^{n} \frac{\varphi_k}{D_{ik}{}^0} \right) = \frac{D_{mix}}{D_{s,i}{}^0} \left( \frac{\varphi_i}{D_{s,n}{}^0} + \sum_{k=1, k \neq i}^{n} \frac{\varphi_k}{D_{s,k}{}^0} \right),$$

(8)

$$B_{ii} = \frac{D_{mix}}{D_{s,i}{}^0} \left( \frac{\varphi_i}{D_{s,n}{}^0} + \frac{1}{D_{mix}} - \frac{\varphi_i}{D_{s,i}{}^0} \right),$$

(9)

$$B_{ij} = \varphi_i \left( \frac{1}{D_{in}{}^0} - \frac{1}{D_{ij}{}^0} \right) = \frac{\varphi_i D_{mix}}{D_{s,i}{}^0} \left( \frac{1}{D_{s,n}{}^0} - \frac{1}{D_{s,j}{}^0} \right),$$

(10)

and finally, we obtain the kinetic equations in the Fickian form

$$\frac{\partial \varphi_i}{\partial t} = \nabla \left[ -\frac{v_0}{RT D_{mix}} C^{-1} \left( D_s{}^0 \nabla \mu_V^{mix} \right) \right], \qquad i = 1 \dots n-1,$$

(11)

where $D_s{}^0$ is the vector of all self-diffusivities. The matrix $C$ is given by

$$\begin{cases} C_{ii} = \dfrac{1}{D_{mix} \varphi_i} + \dfrac{1}{D_{s,n}{}^0} - \dfrac{1}{D_{s,i}{}^0} \\[2mm] C_{ij} = \dfrac{1}{D_{s,n}{}^0} - \dfrac{1}{D_{s,j}{}^0} \end{cases}.$$

## 3. Inversion of the Maxwell-Stefan matrix

To get a tractable expression for the evolution of the volume fraction, the Maxwell-Stefan equations should be inverted. In general, no simple expression can be found in multicomponent systems, but we will prove that the matrix C found with the help of the Darken equation can be analytically inverted which leads to the following final result:

$$\begin{cases} {C_{ii}}^{-1} = \varphi_i D_{s,n}{}^0 D_{mix} \left( \dfrac{1-\varphi_i}{D_{s,n}{}^0} - \dfrac{\varphi_i}{D_{s,i}{}^0} \right) \\ {C_{ij}}^{-1} = \varphi_i \varphi_j D_{s,n}{}^0 D_{mix} \left( \dfrac{1}{D_{s,j}{}^0} - \dfrac{1}{D_{s,n}{}^0} \right) \end{cases}$$

(13)

To obtain Eq. (13), we write the inverse matrix with the help of the comatrix as

$$C^{-1} = \frac{1}{det(C)} \; {}^t com(C).$$

(14)

First, we calculate the determinant of the matrix $C$. Defining for convenience $a_i = \frac{1}{D_{s,n}{}^0} - \frac{1}{D_{s,i}{}^0}$, we have

$$det(C) = \begin{vmatrix} \dfrac{1}{D_{mix}\varphi_1} + a_1 & a_2 & \cdots & a_i & \cdots & a_{n-1} \\ a_1 & \dfrac{1}{D_{mix}\varphi_2} + a_2 & & \cdots & & \\ & & \ddots & \vdots & & \\ \vdots & \vdots & & \dfrac{1}{D_{mix}\varphi_i} + a_i & \cdots & \vdots \\ & & & \vdots & \ddots & \\ a_1 & a_2 & \cdots & a_i & \cdots & \dfrac{1}{D_{mix}\varphi_{n-1}} + a_{n-1} \end{vmatrix}.$$

(15)

Substracting rows $i$ and $i$-$1$ leads to

$$
det(C) = \begin{vmatrix}
\frac{1}{D_{mix}\varphi_1} + a_1 & a_2 & \cdots & a_i & \cdots & a_{n-2} & a_{n-1} \\
-\frac{1}{D_{mix}\varphi_1} & \frac{1}{D_{mix}\varphi_2} & \cdots & 0 & \cdots & 0 & 0 \\
0 & -\frac{1}{D_{mix}\varphi_2} & \ddots & 0 & & & \\
& 0 & \ddots & \frac{1}{D_{mix}\varphi_i} & & \vdots & \vdots \\
\vdots & \vdots & & -\frac{1}{D_{mix}\varphi_i} & \ddots & 0 & \\
& & & 0 & \ddots & \frac{1}{D_{mix}\varphi_{n-2}} & 0 \\
0 & 0 & \cdots & 0 & \cdots & -\frac{1}{D_{mix}\varphi_{n-2}} & \frac{1}{D_{mix}\varphi_{n-1}}
\end{vmatrix}.
$$

$$(16)$$

Then, we develop the determinant according to the first column:

$$
det(C) = \left( \frac{1}{D_{mix}\varphi_1} + a_1 \right)
\begin{vmatrix}
\frac{1}{D_{mix}\varphi_2} & \cdots & 0 & \cdots & 0 & 0 \\
-\frac{1}{D_{mix}\varphi_2} & \ddots & 0 & & & \\
0 & \ddots & \frac{1}{D_{mix}\varphi_i} & & \vdots & \vdots \\
\vdots & & -\frac{1}{D_{mix}\varphi_i} & \ddots & 0 & \\
& & 0 & \ddots & \frac{1}{D_{mix}\varphi_{n-2}} & 0 \\
0 & \cdots & 0 & \cdots & -\frac{1}{D_{mix}\varphi_{n-2}} & \frac{1}{D_{mix}\varphi_{n-1}}
\end{vmatrix}
$$

$$
+ \frac{1}{D_{mix}\varphi_1}
\begin{vmatrix}
a_2 & a_3 & \cdots & a_i & \cdots & a_{n-2} & a_{n-1} \\
-\frac{1}{D_{mix}\varphi_2} & \frac{1}{D_{mix}\varphi_3} & \cdots & 0 & & 0 & 0 \\
0 & -\frac{1}{D_{mix}\varphi_3} & \ddots & & & & \\
& 0 & \ddots & \frac{1}{D_{mix}\varphi_i} & & \vdots & \vdots \\
\vdots & \vdots & & -\frac{1}{D_{mix}\varphi_i} & \ddots & 0 & \\
& & & 0 & \ddots & \frac{1}{D_{mix}\varphi_{n-2}} & 0 \\
0 & 0 & \cdots & 0 & \cdots & -\frac{1}{D_{mix}\varphi_{n-2}} & \frac{1}{D_{mix}\varphi_{n-1}}
\end{vmatrix}
$$

$$(17)$$

The first determinant appearing in this expression is triangular and is thus equal to the product of the diagonal terms. We develop the second determinant according to its first row in order to obtain once again new triangular determinants. This leads after some arithmetic manipulation to

$$det(C) = \left(\frac{1}{D_{mix}\varphi_1} + a_1\right)\prod_{k=2}^{n-1}\frac{1}{D_{mix}\varphi_k}$$
$$+ \frac{1}{D_{mix}\varphi_1}\left(\sum_{j=2}^{n-1}(-1)^j a_j \prod_{k=2,k\neq j}^{j}\left(-\frac{1}{D_{mix}\varphi_k}\right)\prod_{k=i+1}^{n-1}\frac{1}{D_{mix}\varphi_k}\right),$$

(18)

$$det(C) = \left(\frac{1}{D_{mix}\varphi_1} + a_1\right)\prod_{k=2}^{n-1}\frac{1}{D_{mix}\varphi_k} + \frac{1}{D_{mix}\varphi_1}\left(\sum_{j=2}^{n-1}a_j \prod_{k=2,k\neq j}^{n-1}\frac{1}{D_{mix}\varphi_k}\right),$$

(19)

$$det(C) = \prod_{k=1}^{n-1}\frac{1}{D_{mix}\varphi_k} + \left(\sum_{j=1}^{n-1}a_j \prod_{k=1,k\neq j}^{n-1}\frac{1}{D_{mix}\varphi_k}\right),$$

(20)

$$det(C) = \frac{1}{D_{mix}^{n-2}}\left(\prod_{k=1}^{n-1}\frac{1}{\varphi_k}\right)\left(\frac{1}{D_{mix}} + \sum_{j=1}^{n-1}a_j\varphi_j\right).$$

(21)

We then develop the last term in Eq. (21), make use of the definition of $D_{mix}$ in the multicomponent Darken theory and of the mass conservation $\sum_{i=1}^{n}\varphi_i = 1$ in order to obtain

$$\sum_{j=1}^{n-1}a_j\varphi_j = \sum_{j=1}^{n-1}\varphi_j\left(\frac{1}{D_{s,n}^0} - \frac{1}{D_{s,j}^0}\right) = \frac{1}{D_{s,n}^0}\sum_{j=1}^{n-1}\varphi_j - \sum_{j=1}^{n-1}\frac{\varphi_j}{D_{s,j}^0} = \frac{1-\varphi_n}{D_{s,n}^0} - \left(\frac{1}{D_{mix}} - \frac{\varphi_n}{D_{s,n}^0}\right),$$

(22)

$$\sum_{j=1}^{n-1}a_j\varphi_j = \frac{1}{D_{s,n}^0} - \frac{1}{D_{mix}}.$$

(23)

This leads to the final formula for the determinant:

$$det(C) = \frac{1}{D_{mix}^{n-2}D_{s,n}^0}\prod_{k=1}^{n-1}\frac{1}{\varphi_k}$$

(24)

Second, we calculate the diagonal coefficients of the comatrix by substracting columns *i* and *i-1* as before:

$$com(C)_{ii}$$

$$= (-1)^{2i} \begin{vmatrix} \frac{1}{D_{mix}\varphi_1} + a_1 & a_2 & \cdots & a_{i-1} & a_{i+1} & \cdots & a_{n-1} \\ a_1 & \frac{1}{D_{mix}\varphi_2} + a_2 & & \vdots & & \cdots & \\ & & \ddots & \frac{1}{D_{mix}\varphi_{i-1}} + a_{i-1} & & \vdots & \vdots \\ \vdots & \vdots & & & \frac{1}{D_{mix}\varphi_{i+1}} + a_{i+1} & & \vdots \\ & & & \vdots & & \ddots & \\ a_1 & a_2 & \cdots & a_{i-1} & a_{i+1} & \cdots & \frac{1}{D_{mix}\varphi_{n-1}} + a_{n-1} \end{vmatrix}$$

$$(25)$$

$$com(C)_{ii} = \begin{vmatrix} \frac{1}{D_{mix}\varphi_1} + a_1 & a_2 & \cdots & a_{i-1} & a_{i+1} & \cdots & a_{n-2} & a_{n-1} \\ -\frac{1}{D_{mix}\varphi_1} & \frac{1}{D_{mix}\varphi_2} & \cdots & 0 & 0 & \cdots & 0 & 0 \\ 0 & -\frac{1}{D_{mix}\varphi_2} & \ddots & \vdots & \vdots & & & \\ \vdots & & \ddots & \frac{1}{D_{mix}\varphi_{i-1}} & & \cdots & \vdots & \vdots \\ \vdots & & & -\frac{1}{D_{mix}\varphi_{i-1}} & \frac{1}{D_{mix}\varphi_{i+1}} & & \vdots & \vdots \\ \vdots & \vdots & & \vdots & -\frac{1}{D_{mix}\varphi_{i+1}} & \ddots & & \\ & & & & & \ddots & \frac{1}{D_{mix}\varphi_{n-2}} & 0 \\ 0 & 0 & \cdots & 0 & 0 & \cdots & -\frac{1}{D_{mix}\varphi_{n-2}} & \frac{1}{D_{mix}\varphi_{n-1}} \end{vmatrix}$$

$$(26)$$

This can be again expanded using the first column, and then for the $-1/D_{mix}\varphi_1$ term using the first row in a very similar way as compared to the calculation of the determinant. This leads to

$$com(C)_{ii} = \left( \frac{1}{D_{mix}\varphi_1} + a_1 \right) \prod_{k=2, k \neq i}^{n-1} \frac{1}{D_{mix}\varphi_k} + \frac{1}{D_{mix}\varphi_1} \left( \sum_{j=2, j \neq i}^{n-1} a_j \prod_{k=2, k \neq i, j}^{n-1} \frac{1}{D_{mix}\varphi_k} \right),$$

$$(27)$$

$$com(C)_{ii} = \prod_{k=1, k \neq i}^{n-1} \frac{1}{D_{mix}\varphi_k} + \left( \sum_{j=1, j \neq i}^{n-1} a_j \prod_{k=1, k \neq i, j}^{n-1} \frac{1}{D_{mix}\varphi_k} \right),$$

$$(28)$$

$$com(C)_{ii} = \frac{1}{D_{mix}{}^{n-3}} \left( \prod_{k=1, k \neq i}^{n-1} \frac{1}{\varphi_k} \right) \left( \frac{1}{D_{mix}} + \sum_{j=1, j \neq i}^{n-1} a_j \varphi_j \right),$$

(29)

and using again Eq. (23), we obtain

$$\sum_{j=1, j \neq i}^{n-1} a_j \varphi_j = \frac{1}{D_{s,n}{}^0} - \frac{1}{D_{mix}} - a_i \varphi_i = \frac{1 - \varphi_i}{D_{s,n}{}^0} - \frac{1}{D_{mix}} + \frac{\varphi_i}{D_{s,j}{}^0}.$$

(30)

This finally leads to

$$com(C)_{ii} = \frac{1}{D_{mix}{}^{n-3}} \left( \frac{1 - \varphi_i}{D_{s,n}{}^0} + \frac{\varphi_i}{D_{s,j}{}^0} \right) \prod_{k=1, k \neq i}^{n-1} \frac{1}{\varphi_k}.$$

(31)

Third, we calculate the off-diagonal coefficients of the comatrix, i.e. for $i<j$:

(32)

Due to the skipping of the $i^{th}$ line of the initial matrix, the terms with $1/D_{mix}\varphi_k + a_k$ are shifted above the diagonal starting from line $i$, and due to the skipping of the $j^{th}$ column they are back on the diagonal starting from line $j>i$ (with the $j+1$ values instead of the $j$ values). Note that for the off-diagonal coefficients of the comatrix with $j<i$, the $1/D_{mix}\varphi_k + a_k$ terms lie below the diagonal from line $j$ to line $i$ but the derivation is completely similar. The idea is to permute the $i^{th}$ column (where there are no $1/D_{mix}\varphi_k + a_k$ terms) to the last position and then to permute the $j\text{-}1^{th}$ line (where there are no $1/D_{mix}\varphi_k + a_k$ terms) to the last position in order to get

$$com(C)_{ij} = (-1)^{i+j}(-1)^{n-1-i}(-1)^{n-1-(j-1)} \begin{vmatrix} \frac{1}{D_{mix}\varphi_1}+a_1 & a_2 & \cdots & a_{i-1} & a_{i+1} & \cdots & a_{j-1} & a_{j+1} & \cdots & a_{n-2} & a_i \\ a_1 & \frac{1}{D_{mix}\varphi_2}+a_2 & & \vdots & & & & & & & \\ & & \frac{1}{D_{mix}\varphi_{i-1}}+a_{i-1} & & \vdots & & & \vdots & & & \\ & & & \frac{1}{D_{mix}\varphi_{i+1}}+a_{i+1} & & & & \vdots & & & \\ \vdots & \vdots & \vdots & & & \frac{1}{D_{mix}\varphi_{j-1}}+a_{j-1} & & & \vdots & & \\ & & & & & & \frac{1}{D_{mix}\varphi_{j+1}}+a_{j+1} & & \ddots & & \\ & & & & & \vdots & & & \frac{1}{D_{mix}\varphi_{n-1}}+a_{n-1} & & \\ a_1 & a_2 & \cdots & a_{i-1} & a_{i+1} & \cdots & a_{j-1} & a_{j+1} & \cdots & a_{n-1} & a_i \end{vmatrix}.$$

(33)

The prefactor reflects the change of sign for the performed permutations. This determinant is then reduced to a quasi-triangular form with the already described substraction of successive lines:

$$com(C)_{ij}$$
$$= (-1)^{2n-1} \begin{vmatrix} \frac{1}{D_{mix}\varphi_1}+a_1 & a_2 & \cdots & a_{i-1} & a_{i+1} & \cdots & a_{j-1} & a_{j+1} & \cdots & a_{n-2} & a_i \\ -\frac{1}{D_{mix}\varphi_1} & \frac{1}{D_{mix}\varphi_2} & 0 & 0 & 0 & 0 & 0 & 0 \\ 0 & -\frac{1}{D_{mix}\varphi_2} & \ddots & \vdots & \vdots & & & \\ & & \ddots & \frac{1}{D_{mix}\varphi_{i-1}} & 0 & & \vdots & \\ & & & -\frac{1}{D_{mix}\varphi_{i-1}} & \frac{1}{D_{mix}\varphi_{i+1}} & & \vdots & \\ & & & 0 & -\frac{1}{D_{mix}\varphi_{i+1}} & \ddots & & \vdots & \vdots \\ \vdots & \vdots & & & \ddots & \frac{1}{D_{mix}\varphi_{j-1}} & 0 & \vdots \\ & & & & & -\frac{1}{D_{mix}\varphi_{j-1}} & \frac{1}{D_{mix}\varphi_{j+1}} & \\ & & & & & 0 & -\frac{1}{D_{mix}\varphi_{j+1}} & \ddots & \\ & & & & & & & & \frac{1}{D_{mix}\varphi_{n-1}} & 0 \\ 0 & 0 & \cdots & 0 & 0 & \cdots & 0 & 0 & \cdots & -\frac{1}{D_{mix}\varphi_{n-1}} & 0 \end{vmatrix}$$

(34)

Once again, we develop along the first column, and for the second term of the first column we develop along the first line and then obtain triangular determinants with a zero on the diagonal (whose value is thus zero), except for the last term of the development (second term of the first column and then last column). This leads to

$$com(C)_{ij} = (-1)^{2n-1} \frac{1}{D_{mix}\varphi_1} a_i \prod_{k=2, k\neq i,j}^{n-1} \frac{1}{D_{mix}\varphi_k}$$

(35)

and keeping in mind the definition of $a_i$ we get

$$com(C)_{ij} = \frac{1}{D_{mix}{}^{n-3}} \left( \frac{1}{D_{s,i}{}^0} - \frac{1}{D_{s,n}{}^0} \right) \prod_{k=1, k\neq i,j}^{n-1} \frac{1}{\varphi_k}.$$

(36)

Finally, inserting equations (24), (31) and (36) in equation (14) provides the final quite simple expression for the inverse matrix, namely equation (13).

# 4. Final expression of the kinetic equations for Maxwell-Stefan with multicomponent Darken diffusivities

Inserting Eq. (13) into Eq. (11) leads to the Fickian-like expression

$$\frac{\partial \varphi_i}{\partial t} = \nabla \left[ \sum_{j=1}^{n-1} \Lambda_{ij} \nabla \mu_{V,i}^{mix} \right], \qquad i = 1 \dots n-1,$$

(37)

where the mobility coefficients are

$$\begin{cases} \Lambda_{ii} = \dfrac{v_0}{RT} \varphi_i D_{s,n}{}^0 D_{s,i}{}^0 \left( \dfrac{1-\varphi_i}{D_{s,n}{}^0} + \dfrac{\varphi_i}{D_{s,i}{}^0} \right) \\ \Lambda_{ij} = \dfrac{v_0}{RT} \varphi_i \varphi_j D_{s,n}{}^0 D_{s,j}{}^0 \left( \dfrac{1}{D_{s,j}{}^0} - \dfrac{1}{D_{s,n}{}^0} \right) \end{cases}.$$

(38)

Finally, remember that the initial expression of the generalized Maxwell-Stefan expression for polymers has been obtained with the assumption that all segments of the molecules can participate to the friction force, meaning that $D_{s,i}{}^0 = N_i D_{s,i}$. After a final manipulation this leads to

$$\begin{cases} \Lambda_{ii} = \dfrac{v_0}{RT} \varphi_i \big(N_i D_{s,i}(1-\varphi_i) + N_n D_{s,n}\varphi_i\big) \\ \quad \Lambda_{ij} = \dfrac{v_0}{RT} \varphi_i \varphi_j \big(N_n D_{s,n} - N_j D_{s,j}\big) \end{cases}.$$

(39)

# Derivation of an alternative expression for the fast-mode theory

In the fast-mode theory, the time development of the volume fractions is usually written as

$$\frac{\partial \varphi_i}{\partial t} = \nabla \left[ \sum_{j=1}^{n-1} \Lambda_{ij} \nabla \big(\mu_{V,j}^{mix} - \mu_{V,n}^{mix}\big) \right], \qquad i = 1 \dots n-1,$$

(40)

whereby the symmetric mobility coefficients $\Lambda_{ij}$ are found to be

$$\begin{cases} \Lambda_{ii} = (1-\varphi_i)^2 M_i + \varphi_i{}^2 \displaystyle\sum_{k=1, k\neq i}^{n} M_k \\ \Lambda_{ij} = -(1-\varphi_i)\varphi_j M_i - (1-\varphi_j)\varphi_i M_j + \varphi_i \varphi_j \displaystyle\sum_{k=1, k\neq i\neq j}^{n} M_k \end{cases},$$

(41)

with the fluid mobilities

$$M_i = \frac{v_0}{RT} N_i \varphi_i D_{s,i}.$$

(42)

Nevertheless, we can write it in a slightly different form starting from the definition of the flux in the lattice reference frame[38],

$$\begin{cases} \boldsymbol{j}_i = -M_i \nabla(\mu_{V,i}^{mix}) \\ \sum_{i=1}^{n} \boldsymbol{j}_i + \boldsymbol{j}_V = 0 \end{cases},$$

(43)

whereby $\boldsymbol{j}_V$ is the vacancy flux. To ensure incompressibility, the fluxes in the volume reference frame then need to be[38]

$$\boldsymbol{j}_i = -(1-\varphi_i)M_i \nabla\mu_{V,i}^{mix} + \varphi_i \sum_{j=1,j\neq i}^{n} M_j \nabla\mu_{V,j}^{mix} \qquad i = 1 \dots n-1.$$

(44)

We drop the $n^{th}$ term $\nabla\mu_{V,n}^{mix}$ of these equations using the Gibbs-Duhem relationship in the form $\varphi_n \nabla\mu_{V,n}^{mix} = -\sum_{j=1}^{n-1} \varphi_j \nabla\mu_{V,j}^{mix}$,

$$\boldsymbol{j}_i = -(1-\varphi_i)M_i \nabla\mu_{V,i}^{mix} + \varphi_i \sum_{j=1,j\neq i}^{n-1} M_j \nabla\mu_{V,j}^{mix} + \varphi_i M_n \nabla\mu_{V,n}^{mix},$$

(45)

$$\boldsymbol{j}_i = -(1-\varphi_i)M_i \nabla\mu_{V,i}^{mix} + \varphi_i \sum_{j=1,j\neq i}^{n-1} M_j \nabla\mu_{V,j}^{mix} - \frac{\varphi_i M_n}{\varphi_n} \sum_{j=1}^{n-1} \varphi_j \nabla\mu_{V,j}^{mix},$$

(46)

$$\boldsymbol{j}_i = -(1-\varphi_i)M_i \nabla\mu_{V,i}^{mix} - \frac{\varphi_i^2 M_n}{\varphi_n} \nabla\mu_{V,i}^{mix} + \sum_{j=1,j\neq i}^{n-1} \varphi_i M_j \nabla\mu_{V,j}^{mix} - \sum_{j=1,j\neq i}^{n-1} \frac{\varphi_i \varphi_j M_n}{\varphi_n} \nabla\mu_{V,j}^{mix}.$$

(47)

Grouping both sums together and writing them in the form of Eq. (37) we find the mobility coefficients to be

$$\begin{cases} \Lambda_{ii} = (1-\varphi_i)M_i + \dfrac{\varphi_i^2 M_n}{\varphi_n} \\ \Lambda_{ij} = -\varphi_i M_j + \dfrac{\varphi_i \varphi_j M_n}{\varphi_n} \end{cases}.$$

(48)

Reminding Eq.(42), this can be written as

$$\begin{cases} \Lambda_{ii} = \dfrac{v_0}{RT} \varphi_i \Big( (1 - \varphi_i) N_i D_{s,i} + \varphi_i N_n D_{s,n} \Big) \\ \qquad \Lambda_{ij} = \dfrac{v_0}{RT} \varphi_i \varphi_j \big( N_n D_{s,n} - N_j D_{s,j} \big) \end{cases},$$

$$(49)$$

which is exactly Eq. (39) and hence proves the equivalence of both theories.

# Conclusion

The fast-mode theory seems to be recognized in the polymer physics community as one of the most satisfactory theories to describe diffusion and spinodal decomposition within the assumptions of continuum mechanics. Describing the mass transport in miscible systems by the Maxwell-Stefan theory, the chemical engineering community pointed out similar advantages for the multicomponent Darken equation to estimate diffusivities. We proved analytically the strict equivalence between both theories for multicomponent mixtures, provided the Darken equation is used for the expression of the Maxwell-Stefan diffusivities. In this paper, the driving force only depends on the chemical potential gradients, but the derivation is still valid with additional or different driving forces without loss of generality. The equivalence holds even for molecules of very dissimilar sizes with the important restriction that the friction of the whole molecule is the sum of the friction of all its equally contributing segments. Note that in the case of polymers, this corresponds only to the situation of unentangled dynamics with the hypothesis of a Rouse behavior of the chain. Here, we point out that the argument of neglecting cross-correlations of velocities of different molecules has been used in both, the derivation of the multicomponent Darken equation and in the derivation of the two-component fast-mode equation by Akcasu [42]. From the equivalence shown in this paper, we can propose that negligible cross-correlations are a main underlying hypothesis of the multicomponent fast-mode theory as well. In general, the complexity of the diffusive behavior of the macromolecule is taken into account neither in the Maxwell-Stefan and Darken theory, nor in the fast-mode theory. Not only the volume-fraction dependency of the diffusion coefficients is questionable, but also the applicability of the basic kinetic equations, since both rely on the assumption of no cross correlations for the velocity of a different molecule/segment.

Our findings allow to reduce the choice of a suitable theory for investigating multicomponent systems: both most successful theories for respectively immiscible and miscible systems have been unified. Thereby, the fast-mode form of the equations is certainly the most appropriate for code implementation in simulation studies since these equations are much more tractable.

# Acknowledgements

The authors acknowledge financial support by the Deutsche Forschungsgemeinschaft (DFG) within the Cluster of Excellence "Engineering of Advanced Materials" (project EXC 315) (Bridge Funding).

# Author information

Corresponding Author

*E-mail: (J.H.) mailto:j.harting@fz-juelich.de